\documentclass[final,5p,times,twocolumn]{elsarticle}
\usepackage{lineno,hyperref,amsmath,amsthm,amssymb,amsfonts,ragged2e,color,subfig}
\journal{``AIP Advances''}
\begin{document}
\begin{frontmatter}
\title{Ion-acoustic shock waves in a magnetized plasma featuring super-thermal distribution}
\author{N.M. Heera$^{*,1}$, J. Akter$^{**,1}$, N.K. Tamanna$^{***,1}$, N.A. Chowdhury$^{\dag,2}$,
T.I. Rajib$^{\ddag,1}$, S. Sultana$^{\S,1}$, and A.A. Mamun$^{\S\S,1,3}$}
\address{$^{1}$Department of Physics, Jahangirnagar University, Savar, Dhaka-1342, Bangladesh\\
$^{2}$Plasma Physics Division, Atomic Energy Centre, Dhaka-1000, Bangladesh\\
$^{3}$Wazed Miah Science Research Centre, Jahangirnagar University, Savar, Dhaka-1342, Bangladesh\\
e-mail: $^*$heera112phys@gmail.com, $^{**}$akter277phy@gmail.com, $^{***}$tamanna1995phy@gmail.com,\\
$^{\dag}$nurealam1743phy@gmail.com, $^{\ddag}$tirajibphys@gmail.com, $^{\S}$ssultana@juniv.edu, $^{\S\S}$mamun\_phys@juniv.edu}
\begin{abstract}
A theoretical investigation has been made on the propagation of ion-acoustic (IA) shock waves (IASHWs) in
a magnetized pair-ion plasma having inertial warm positive and negative ions, and inertialess super-thermal
electrons and positrons. The well known Burgers' equation has been derived  by employing the reductive
perturbation method. The plasma model supports both positive and negative shock structures
under consideration of super-thermal electrons and positrons. It is found that  the oblique angle ($\delta$)
enhances the magnitude of the amplitude of both positive and negative shock profiles.
It is also observed that the steepness of the shock profiles decreases with the kinematic viscosity of the ion,
and the height of the shock profile increases (decreases) with the mass of the positive (negative) ion.
The implications of the results have been briefly discussed for space and laboratory plasmas.
\end{abstract}
\end{frontmatter}
\section{Introduction}
\label{1sec:Introduction}
The negative ions have been identified in cometary comae \cite{Chaizy1991}, ($H^+$, $O_2^-$) and ($H^+$, $H^-$) plasmas in the D
and F-regions of Earth's ionosphere \cite{Massey1976}, upper regions of Titan's atmosphere \cite{Massey1976,Coates2007},
and also in the laboratory experiments, namely, plasma processing reactors \cite{Gottscho1986},  ($Ar^+$, $SF_6^-$) plasma \cite{Wong1975,Nakamura1997,Cooney1991,Nakamura1999}, ($Ar^+$, $O_2^-$) plasma, ($K^+$, $SF_6^-$) plasma \cite{Song1991,Sato1994},
($Xe^+$, $F^-$) plasma \cite{Ichiki2002}, neutral beam sources \cite{Bacal1979}, plasma etching \cite{Sheehan1988},
($Ar^+$, $F^-$) plasma \cite{Nakamura1984}, combustion products \cite{Sheehan1988},
and Fullerene ($C_{60}^+$, $C_{60}^-$) plasma \cite{Oohara2003,Oohara2005}, etc. The negative ion in a plasma is considered
to be produced due to the attachment of electrons with the atom \cite{Shukla2002}.
The existence of positron \cite{Zhang2005,Sturrock1971,Harding1998,Harding2006} in the astro-physical environments
not only changes the dynamics of the plasma system but also changes the configuration of the nonlinear electrostatic structures \cite{Ahmed2018,Esfandyari-Kalejahi2006,C1}. Ahmed \textit{et al.} \cite{Ahmed2018} examined the stability of
ion-acoustic (IA) waves (IAW) in a pair-ion (PI) plasma (PIP) medium (PIPM) in the presence of positrons,
and observed that the height of the electrostatic waves decreases with an increase in the value of positron concentration.
Esfandyari-Kalejahi \textit{et al.} \cite{Esfandyari-Kalejahi2006} studied obliquely modulated electrostatic
modes and envelope excitations in a PIPM having electrons and positrons.

Vasylinaus \cite{Vasyliunas1968} first demonstrated the super-thermal $\kappa$-distribution for explaining
the high energy tails which can be observed in the non-equilibrium plasma systems \cite{Kaladze2014,Pakzad2011,Bansal2020,C2},
and introduced the parameter $\kappa$ in the super-thermal $\kappa$-distribution for measuring the amount of
deviation of the plasma species from Maxwell-Boltzmann distributions.
Kaladze and Mahmood \cite{Kaladze2014} investigated the nonlinear features of IAWs in a super-thermal plasma,
and observed that the super-thermal electrons can cause to decrease the height of the IAWs.
Pakzad \cite{Pakzad2011} considered a three-component plasma model having inertial ions, and inertialess
super-thermal electrons and positrons, and examined the IA shock waves (IASHWs), and reported that
the amplitude of the shock profile increases with $\kappa$. Bansal \textit{et al.} \cite{Bansal2020}
studied cylindrical and spherical IASHWs in a dusty plasma having super-thermal electrons.

The energy dissipation of the electrostatic wave, which is governed by the Burgers' equation \cite{Burgers1948},
may arise due to the kinematic viscosity of the medium \cite{Abdelwahed2016,El-Hanbaly2016,Ghai2017}.
Abdelwahed \textit{et al.} \cite{Abdelwahed2016} analyzed IASHWs in a PIP, and found that the steepness
of the shock potential decreases with increasing
ion viscosity. El-Hanbaly \textit{et al.} \cite{El-Hanbaly2016} studied dust-acoustic shock waves in a dusty
plasma having super-thermal electrons, and observed that the steepness of the shock profile decreases but the amplitude remains constant
with the variation of dust kinematic viscosity, and also reported that the height of the positive shock
potential increases with $\kappa$. Ghai and Saini \cite{Ghai2017} examined that the amplitude of the electrostatic
shock structure enhances with the super-thermality of the plasma species in a dusty plasma medium.

The external magnetic field can significantly change the shape of the nonlinear electrostatic excitations
\cite{Hossen2017,Bains2015,Bansal2019}. Hossen \textit{et al.} \cite{Hossen2017} observed that the magnitude of the amplitude of
positive and negative electrostatic shock structures increases with the oblique angle ($\delta$) which
arises due to the external magnetic field. Bains \textit{et al.} \cite{Bains2015} considered a two-component plasma medium containing inertial
ions and inertialess super-thermal electrons, and observed IASHWs in the presence of external magnetic field,
and found that the magnitude of the amplitude of negative shock potential increases with $\kappa$.
Bansal \textit{et al.} \cite{Bansal2019} examined the effects of positron density and temperature on the
electron-acoustic shock waves in a magnetized dissipative plasma. To the best knowledge of the authors, no attempt
has been made to study the IASHWs in a four-component magnetized plasma by considering kinematic viscosities
of both inertial warm positive and negative ion species, and inertialess super-thermal electrons and positrons. The aim of our present
investigation is, therefore, to derive Burgers' equation, and will investigate the IASHWs in a four-component magnetized PIP,
and to observe the effects of various plasma parameters on the configuration of IASHWs.

The outline of the paper is as follows: The basic equations are displayed in section \ref{1sec:Governing equations}.
The Burgers equation has been derived in section \ref{1sec:Derivation of the Burgers equation}.
Results and discussion are reported in section \ref{1sec:Results and discussion}.
A brief conclusion is provided in section \ref{1sec:Conclusion}.
\section{Governing equations}
\label{1sec:Governing equations}
We consider a magnetized PIPM consisting of inertial positively
charged warm ions (mass $m_+$; charge $eZ_+$; temperature $T_+$; number density $\tilde{n}_{+}$),
negatively charged warm ions (mass $m_-$; charge $-eZ_-$; temperature $T_-$; number density $\tilde{n}_{-}$),
inertialess electrons (mass $m_e$; charge $-e$; temperature $T_e$; number density $\tilde{n}_e$),
and inertialess positrons (mass $m_p$; charge $e$; temperature $T_p$; number density $\tilde{n}_p$);
where $Z_+$ ($Z_-$) is the charge state of the positive (negative) ion, and $e$ is the magnitude of the charge of an electron.
Thus, the quasi-neutrality condition at equilibrium for our considered plasma medium reads: $n_{e0}+Z_-n_{-0}\approx n_{p0}+Z_+n_{+0}$,
where $n_{s0}$ is the unperturbed number densities of plasma species $s$ and $s=-,\,+,\,e,\,p$ refer to the negative ions,
positive ions, electrons, and positrons, respectively.
An external magnetic field $\mbox{B}$ has been considered in the system directed along
the $z$-axis defining $\mbox{B}=\mbox{B}_0\hat{z}$, where $\mbox{B}_0$ and $\hat{z}$ are the strength
of the external magnetic field and unit vector directed along the $z$-axis, respectively.
The dynamics of the magnetized PIP in the presence of super-thermal electrons and positrons is governed by the following
set of equations \cite{Adhikary2012,Dev2018,Atteyaa2018}
\begin{eqnarray}
&&\hspace*{-1.3cm}\frac{\partial \tilde{n}_{+}}{\partial \tilde{t}}+\acute{\nabla}\cdot(\tilde{n}_+ \tilde{u}_+)=0,
\label{1eq:1}\\
&&\hspace*{-1.3cm}\frac{\partial \tilde{u}_+}{\partial\tilde{t}}+(\tilde{u}_+\cdot\acute{\nabla})\tilde{u}_{+}=-\frac{Z_+e}{m_+}\acute{\nabla}\tilde{\psi} +\frac{Z_+eB_0}{m_+}(\tilde{u}_{+}\times\hat{z})
\nonumber\\
&&\hspace*{1.5cm}-\frac{1}{m_+\tilde{n}_+}\acute{\nabla} P_+ +\tilde{\eta}_+\acute{\nabla}^2\tilde{u}_+,
\label{1eq:2}\\
&&\hspace*{-1.3cm}\frac{\partial\tilde{n}_-}{\partial \tilde{t}}+\acute{\nabla}\cdot(\tilde{n}_-\tilde{u}_-)=0,
\label{1eq:3}\\
&&\hspace*{-1.3cm}\frac{\partial\tilde{u}_-}{\partial \tilde{t}}+(\tilde{u}_-\cdot\acute{\nabla})\tilde{u}_{-}=\frac{Z_-e}{m_-}\acute{\nabla}\tilde{\psi} -\frac{Z_-eB_0}{m_-}(\tilde{u}_-\times\hat{z})
\nonumber\\
&&\hspace*{1.5cm}-\frac{1}{m_-\tilde{n}_-}\acute{\nabla} P_- +\tilde{\eta}_-\acute{\nabla}^2\tilde{u}_-,
\label{1eq:4}\\
&&\hspace*{-1.3cm}\acute{\nabla}^2\tilde{\psi}=4\pi e[\tilde{n}_e-\tilde{n}_p+Z_-\tilde{n}_--Z_+\tilde{n}_+],
\label{1eq:5}\
\end{eqnarray}
where $\tilde{u}_+$ ($\tilde{u}_-$) is the positive (negative)
ion fluid speed, $\tilde{\eta}_+=\mu/m_+\tilde{n}_{+}$ ($\tilde{\eta}_-=\mu/m_-\tilde{n}_{-}$) is the kinematic viscosity of the positive (negative)
ion, $P_+$ ($P_-$) is the pressure of positive (negative) ion, and $\tilde{\psi}$ represents the electrostatic wave potential.
Now, we  are introducing normalized parameters, namely, $n_+\rightarrow\tilde{n}_+/n_{+0}$, $n_-\rightarrow\tilde{n}_-/n_{-0}$, $n_e\rightarrow\tilde{n}_e/n_{e0}$, and $n_p\rightarrow\tilde{n}_p/n_{p0}$,
$u_+\rightarrow\tilde{u}_+/C_-$, $u_-\rightarrow\tilde{u}_-/C_-$ [where $C_-=(Z_-k_BT_e/m_-)^{1/2}$ with $k_B$ being the Boltzmann constant]; $\psi\rightarrow\tilde{\psi}e/k_BT_e$; $t=\tilde{t}/\omega_{P_-}^{-1}$ [where $\omega_{P_-}^{-1}=(m_-/4\pi e^{2}Z_-^{2}n_{-0})^{1/2}$]; $\nabla=\acute{\nabla}/\lambda_{D}$ [where $\lambda_{D}=(k_BT_e/4\pi e^2Z_-n_{-0})^{1/2}$]. The pressure term of the positive and negative ions can be recognized as $P_{\pm}=P_{\pm0}(\tilde{n}_\pm/n_{\pm0})^\gamma$ with $P_{\pm0}=n_{\pm0}k_BT_\pm$ being the equilibrium
pressure of the positive (for $+0$ sign) and negative (for $-0$ sign) ions, and
$\gamma=(N+2)/N$ (where $N$ is the degree of freedom and for three-dimensional case
$N=3$, then $\gamma=5/3$). For simplicity, we have considered ($\eta=\tilde{\eta}_+\approx\tilde{\eta}_-$), and $\eta$ is
normalized by $\omega_{p_-}\lambda_D^{2}$.
By using the normalizing factors mentioned above, one can obtain the normalized form of
Eqs. \eqref{1eq:1}$-$\eqref{1eq:5} as follows:
\begin{eqnarray}
&&\hspace*{-1.3cm}\frac{\partial n_+}{\partial t}+\nabla\cdot(n_+u_+)=0,
\label{1eq:6}\\
&&\hspace*{-1.3cm}\frac{\partial u_+}{\partial t}+(u_+\cdot\nabla)u_+=-\sigma_1\nabla\psi+\sigma_1\Omega_c(u_+\times\hat{z})
\nonumber\\
&&\hspace*{1.5cm}-\sigma_2\nabla n_+^{\gamma-1}+\eta\nabla^2u_+,
\label{1eq:7}\\
&&\hspace*{-1.3cm}\frac{\partial n_-}{\partial t}+\nabla\cdot(n_-u_-)=0,
\label{1eq:8}\\
&&\hspace*{-1.3cm}\frac{\partial u_-}{\partial t}+(u_-\cdot\nabla)u_-=\nabla\psi-\Omega_c(u_-\times\hat{z})
\nonumber\\
&&\hspace*{1.5cm}-\sigma_3\nabla n_-^{\gamma-1}+\eta\nabla^2u_-,
\label{1eq:9}\\
&&\hspace*{-1.3cm}\nabla^2\psi=\mu_en_e-\mu_pn_p-(1+\mu_e-\mu_p)n_++n_-.
\label{1eq:10}\
\end{eqnarray}
Other plasma parameters are considered as $\sigma_1=Z_+m_-/Z_-m_+$,
$\sigma_2=\gamma T_+m_-/[(\gamma-1)Z_-T_em_+]$,
$\sigma_3=\gamma T_-/[(\gamma-1)Z_-T_e]$, $\mu_e = n_{e0}/Z_-n_{-0}$, $\mu_p = n_{p0}/Z_-n_{-0}$, and $\Omega_c=\omega_c/\omega_{p_-}$ [where $\omega_c=Z_-eB_0/m_-$].
The expressions for the number densities of the super-thermal electrons and positrons (following the $\kappa$-distribution) can
be represented as, respectively, \cite{Kaladze2014,Pakzad2011,Bansal2020}
\begin{eqnarray}
&&\hspace*{-1.3cm}n_e=\bigg[1-\frac{\psi}{(\kappa_e-3/2)}\bigg]^{-\kappa_e+\frac{1}{2}},
\label{1eq:11}\\
&&\hspace*{-1.3cm}n_p=\bigg[1+\frac{\sigma_4\psi}{(\kappa_p-3/2)}\bigg]^{-\kappa_p+\frac{1}{2}},
\label{1eq:12}\
\end{eqnarray}
where $\kappa_e$ and $\kappa_p$ are the spectral indices of super-thermal electrons and positrons, respectively,
and $\sigma_4=T_e/T_p$. It may be noted here that for our numerical analysis, we have considered $\kappa=\kappa_e=\kappa_p$.
The super-thermal $\kappa$-distribution is meaningless for $\kappa<3/2$, and the $\kappa$-distribution behaves like
the Maxwell-Boltzmann distribution for large values of $\kappa\rightarrow\infty$.
Now, by substituting Eqs. \eqref{1eq:11} and \eqref{1eq:12} into the Eq. \eqref{1eq:10}, and expanding up to third order in $\psi$, we get
\begin{eqnarray}
&&\hspace*{-1.3cm}\nabla^2 \psi=\mu_e-\mu_p-(1+\mu_e-\mu_p)n_++n_-
\nonumber\\
&&\hspace*{0.0cm}+\mu_1\psi+\mu_2\psi^2+\mu_3\psi^3+\cdot\cdot\cdot,
\label{1eq:13}\
\end{eqnarray}
where the appearance of $\mu_1,\,\mu_2$, and $\mu_3$ in Eq. \eqref{1eq:13}
introduces the super-thermality effect of electrons and positrons in our
considered plasma medium, and is defined as
\begin{eqnarray}
&&\hspace*{-1.3cm}\mu_1= \frac{[(\mu_e+\mu_p\sigma_4)(2\kappa-1)]}{(2\kappa-3)},
\nonumber\\
&&\hspace*{-1.3cm}\mu_2=\frac{[(\mu_e-\mu_p\sigma_4^2)(2\kappa-1)(2\kappa+1)]}{2(2\kappa-3)^2},
\nonumber\\
&&\hspace*{-1.3cm}\mu_3=\frac{[(\mu_e+\mu_p\sigma_4^3)(2\kappa-1)(2\kappa+1)(2\kappa+3)]}{6(2\kappa-3)^3}.
\nonumber\
\end{eqnarray}
\section{Derivation of the Burgers' equation}
\label{1sec:Derivation of the Burgers equation}
To derive the Burgers' equation for the IASHWs propagating in a magnetized PIP,
first we consider reductive perturbation method \cite{C3}, and introduce the
stretched co-ordinates \cite{Hossen2017,Washimi1966}
\begin{eqnarray}
&&\hspace*{-1.3cm}\xi=\epsilon(l_xx+l_yy+l_zz-\Upsilon_p t),
\label{1eq:14}\\
&&\hspace*{-1.3cm}\tau={\epsilon}^2 t,
\label{1eq:15}\
\end{eqnarray}
where $\Upsilon_p$ is the phase speed and $\epsilon$ is a smallness parameter measuring the weakness of
the dissipation ($0<\epsilon<1$). The $l_x$, $l_y$, and $l_z$ (i.e., $l_x^2+l_y^2+l_z^2=1$) are
the directional cosines of the wave vector $k$ along $x$, $y$, and $z$-axes, respectively. Then,
the dependent variables can be expressed in power series of $\epsilon$ as \cite{Hossen2017}
\begin{eqnarray}
&&\hspace*{-1.3cm}n_{+}=1+\epsilon n_{+}^{(1)}+\epsilon^2 n_{+}^{(2)}+\epsilon^3 n_{+}^{(3)}+\cdot\cdot\cdot,
\label{1eq:16}\\
&&\hspace*{-1.3cm}n_{-}=1+\epsilon n_{-}^{(1)}+\epsilon^2 n_{-}^{(2)}+\epsilon^3 n_{-}^{(3)}+\cdot\cdot\cdot,
\label{1eq:17}\\
&&\hspace*{-1.3cm}u_{+x,y}=\epsilon^2 u_{+x,y}^{(1)}+\epsilon^3 u_{+x,y}^{(2)}+\cdot\cdot\cdot,
\label{1eq:18}\\
&&\hspace*{-1.3cm}u_{-x,y}=\epsilon^2 u_{-x,y}^{(1)}+\epsilon^3 u_{-x,y}^{(2)}+\cdot\cdot\cdot,
\label{1eq:19}\\
&&\hspace*{-1.3cm}u_{+z}=\epsilon u_{+z}^{(1)}+\epsilon^2 u_{+z}^{(2)}+\cdot\cdot\cdot,
\label{1eq:20}\\
&&\hspace*{-1.3cm}u_{-z}=\epsilon u_{-z}^{(1)}+\epsilon^2 u_{-z}^{(2)}+\cdot\cdot\cdot,
\label{1eq:21}\\
&&\hspace*{-1.3cm}\psi=\epsilon \psi^{(1)}+\epsilon^2\psi^{(2)}+\cdot\cdot\cdot.
\label{1eq:22}\
\end{eqnarray}
Now, by substituting Eqs. \eqref{1eq:14}$-$\eqref{1eq:22} into Eqs. \eqref{1eq:6}$-$\eqref{1eq:9}, and
\eqref{1eq:13}, and collecting the terms containing $\epsilon$, the first-order equations are reduced to
\begin{eqnarray}
&&\hspace*{-1.3cm} n_{+}^{(1)}=\frac{3\sigma_1l_z^2}{3\Upsilon_p^2-2\sigma_2l_z^2}\psi^{(1)},
\label{1eq:23}\\
&&\hspace*{-1.3cm}u_{+z}^{(1)}=\frac{3\Upsilon_p\sigma_1l_z}{3\Upsilon_p^2-2\sigma_2l_z^2}\psi^{(1)},
\label{1eq:24}\\
&&\hspace*{-1.3cm}n_{-}^{(1)}=-\frac{3l_z^2}{3\Upsilon_p^2-2\sigma_3l_z^2}\psi^{(1)},
\label{1eq:25}\\
&&\hspace*{-1.3cm}u_{-z}^{(1)}=-\frac{3\Upsilon_pl_z}{3\Upsilon_p^2-2\sigma_3l_z^2}\psi^{(1)}.
\label{1eq:26}\
\end{eqnarray}
Now, by using the expression of $n_{+}^{(1)}$, $u_{+z}^{(1)}$, $n_{-}^{(1)}$, $u_{-z}^{(1)}$ from
 Eqs. \eqref{1eq:23}$-$\eqref{1eq:26}, the phase speed of IASHWs can be written as
\begin{eqnarray}
&&\hspace*{-1.3cm}\Upsilon_{p}\equiv {\Upsilon_{p+}}=l_z\sqrt{{\frac{-b_1+\sqrt{b_1^2-36\sigma_1b_2}}{18\sigma_1}}},
\label{1eq:27}\\
&&\hspace*{-1.3cm}\Upsilon_{p}\equiv {\Upsilon_{p-}}=l_z\sqrt{{\frac{-b_1-\sqrt{b_1^2-36\sigma_1b_2}}{18\sigma_1}}},
\label{1eq:28}\
\end{eqnarray}
where $b_1=9\sigma_1\mu_p-6\sigma_3\mu_1-6\sigma_2\mu_1-9-9\sigma_1\mu_e-9\sigma_1$
and $b_2=4\sigma_2\sigma_3\mu_1+6\sigma_2+6\sigma_1\sigma_3\mu_e-6\sigma_1\sigma_3\mu_p+6\sigma_1\sigma_3$. Equations
\eqref{1eq:27} and \eqref{1eq:28} suggest that our considered plasma medium supports two electrostatic modes,  fast mode- described by
the phase speed in (\ref{1eq:27}) and slow wave- represented by the phase speed in (\ref{1eq:28}), and both waves are propagated obliquely
to the external magnetic field.
The $x$ and $y$-components of the first-order momentum equations can be presented as
\begin{eqnarray}
&&\hspace*{-1.3cm}u_{+x}^{(1)}=-\frac{3l_y\Upsilon_p^2}{\Omega_{c}(3\Upsilon_p^2-2\sigma_2l_z^2)}~\frac{\partial\psi^{(1)}}{\partial\xi},
\label{1eq:29}\\
&&\hspace*{-1.3cm}u_{+y}^{(1)}=\frac{3l_x\Upsilon_p^2}{\Omega_{c}(3\Upsilon_p^2-2\sigma_2l_z^2)}~\frac{\partial\psi^{(1)}}{\partial\xi},
\label{1eq:30}\\
&&\hspace*{-1.3cm}u_{-x}^{(1)}=-\frac{3l_y\Upsilon_p^2}{\Omega_{c}(3\Upsilon_p^2-2\sigma_3l_z^2)}~\frac{\partial\psi^{(1)}}{\partial\xi},
\label{1eq:31}\\
&&\hspace*{-1.3cm} u_{-y}^{(1)}=\frac{3l_x\Upsilon_p^2}{\Omega_{c}(3\Upsilon_p^2-2\sigma_3l_z^2)}~\frac{\partial \psi^{(1)}}{\partial\xi}.
\label{1eq:32}\
\end{eqnarray}
Now, by taking the next higher-order terms, the equation of continuity, momentum equation, and Poisson's equation
can be written as
\begin{eqnarray}
&&\hspace*{-1.3cm}\frac{\partial n_{+}^{(1)}}{\partial\tau}-\Upsilon_p\frac{\partial n_{+}^{(2)}}{\partial\xi}+l_x\frac{\partial u_{+x}^{(1)}}{\partial\xi}+l_y\frac{\partial u_{+y}^{(1)}}{\partial\xi}
\nonumber\\
&&\hspace*{1.0cm}+l_z\frac{\partial u_{+z}^{(2)}}{\partial\xi}+l_z\frac{\partial}{\partial\xi}\big(n_{+}^{(1)}u_{+z}^{(1)}\big)=0,
\label{1eq:33}\\
&&\hspace*{-1.3cm}\frac{\partial u_{+z}^{(1)}}{\partial\tau}-{\Upsilon_p}\frac{\partial u_{+z}^{(2)}}{\partial\xi}+l_zu_{+z}^{(1)}\frac{\partial u_{+z}^{(1)}}{\partial\xi}+\sigma_1l_z\frac{\partial\psi^{(2)}}{\partial\xi}
\nonumber\\
&&\hspace*{-0.2cm}+\frac{2\sigma_2l_z}{3}\frac{\partial n_{+}^{(2)}}{\partial\xi}-\frac{\sigma_2 l_z }{9}\frac{\partial (n_{+}^{(1)})^2}{\partial\xi}-\eta\frac{\partial^2u_{+z}^{(1)}}{\partial\xi^2}=0,
\label{1eq:34}\\
&&\hspace*{-1.3cm}\frac{\partial n_{-}^{(1)}}{\partial\tau}-\Upsilon_p\frac{\partial n_{-}^{(2)}}{\partial\xi}+l_x\frac{\partial u_{-x}^{(1)}}{\partial\xi}+l_y\frac{\partial u_{-y}^{(1)}}{\partial\xi}
\nonumber\\
&&\hspace*{1.2cm}+l_z\frac{\partial u_{-z}^{(2)}}{\partial\xi}+l_z\frac{\partial}{\partial\xi}\big(n_{-}^{(1)}u_{-z}^{(1)}\big)=0,
\label{1eq:35}\\
&&\hspace*{-1.3cm}\frac{\partial u_{-z}^{(1)}}{\partial\tau}-{\Upsilon_p}\frac{\partial u_{-z}^{(2)}}{\partial\xi}+l_zu_{-z}^{(1)}\frac{\partial u_{-z}^{(1)}}{\partial\xi}-l_z\frac{\partial\psi^{(2)}}{\partial\xi}
\nonumber\\
&&\hspace*{-0.1cm}+\frac{2\sigma_3}{3}l_z\frac{\partial n_{-}^{(2)} }{\partial\xi}-\frac{\sigma_3 l_z}{9}\frac{\partial (n_{-}^{(1)})^2 }{\partial\xi}-\eta\frac{\partial^2u_{-z}^{(1)}}{\partial\xi^2}=0,
\label{1eq:36}\\
&&\hspace*{-1.3cm}\sigma_1\psi^{(2)}+\sigma_2{[\psi^{(1)}]}^2+n_-^{(2)}-(1+\mu_e-\mu_p)n_+^{(2)}=0.
\label{1eq:37}\
\end{eqnarray}
Finally, the next higher-order terms of Eqs. \eqref{1eq:6}$-$\eqref{1eq:9},
and \eqref{1eq:13}, with the help of Eqs. \eqref{1eq:23}$-$\eqref{1eq:37},
can provide the Burgers' equation which can be written as
\begin{eqnarray}
&&\hspace*{-1.3cm}\frac{\partial\Psi}{\partial\tau}+A\Psi\frac{\partial\Psi}{\partial\xi}=C\frac{\partial^2\Psi}{\partial\xi^2},
\label{1eq:38}\
\end{eqnarray}
where $\Psi=\psi^{(1)}$ is used for simplicity. In Eq. \eqref{1eq:38}, the nonlinear coefficient $A$
and dissipative coefficient $C$ are given by
\begin{eqnarray}
&&\hspace*{-1.3cm}A=\frac{81\sigma_1^2v_p^2p_1^3l_z^4+F_1}{18v_p p_1l_z^2p_2^3+F_2},~~~\mbox{and}~~~C= \frac{\eta}{2},
\label{1eq:39}\
\end{eqnarray}
where
\begin{eqnarray}
&&\hspace*{-1.3cm}F_1 =81\mu_e\sigma_1^2\Upsilon_p^2p_1^3l_z^4-81\Upsilon_p^2p_1^3l_z^4+2\mu_e\sigma_2\sigma_1^2p_1^3l_z^6
\nonumber\\
&&\hspace*{-0.5cm}+2\sigma_2\sigma_1^2p_1^3l_z^6+2\sigma_3p_1^3l_z^6-2\sigma_2p_1^3p_2^3,
\nonumber\\
&&\hspace*{-1.3cm}F_2=18\sigma_1\Upsilon_p p_2l_z^2p_1^3+18\sigma_1\mu_e\Upsilon_pp_2l_z^2p_1^3 ,
\nonumber\\
&&\hspace*{-1.3cm}p_1=3\Upsilon_p^2-2\sigma_3l_z^2,~~~p_2 = 3\Upsilon_p^2 - 2\sigma_2l_z^2 .
\nonumber\
\end{eqnarray}
\begin{figure}
\centering
\includegraphics[width=80mm]{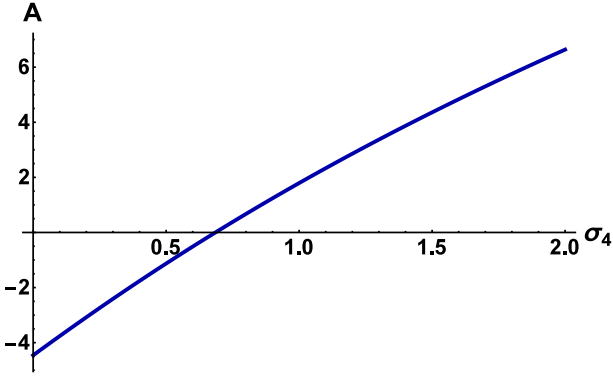}
\caption{The variation of nonlinear coefficient $A$ with $\sigma_4$ when $\delta= 30^\circ$, $\eta=0.3$,
$\kappa=1.6$, $\mu_e=0.5$, $\mu_p=0.3$, $\sigma_1=1.2$, $\sigma_2=0.2$, $\sigma_3=0.02$, and $\Upsilon_p=\Upsilon_{p+}$.}
 \label{1Fig:F1}
\end{figure}
Now, we look for stationary shock wave solution of this Burgers' equation \eqref{1eq:38} by
considering $\zeta =\xi-U_0\tau'$ and $\tau =\tau'$ (where $U_0$ is the speed of the shock waves in the reference frame).
These allow us to write the stationary shock wave solution as \cite{Hossen2017,Karpman1975,Hasegawa1975}
\begin{eqnarray}
&&\hspace*{-1.3cm}\Psi=\Psi_m\Big[1-\tanh\bigg(\frac{\zeta}{\Delta}\bigg)\Big],
\label{1eq:40}\
\end{eqnarray}
where the amplitude $\Psi_m$ and width $\Delta$ are, respectively, given by
\begin{eqnarray}
&&\hspace*{-1.3cm}\Psi_m=\frac{U_0}{A},~~~~\mbox{and}~~~~\Delta=\frac{2C}{U_0}.
\label{1eq:41}\
\end{eqnarray}
It is clear from  Eqs. \eqref{1eq:41} that the IASHWs exist,
which are formed due to the balance between nonlinearity and dissipation,
because $C>0$ and the IASHWs with $\Psi>0$ ($\Psi<0$) exist if $A>0$ ($A<0$) because $U_0>0$.
\begin{figure}
\centering
\includegraphics[width=80mm]{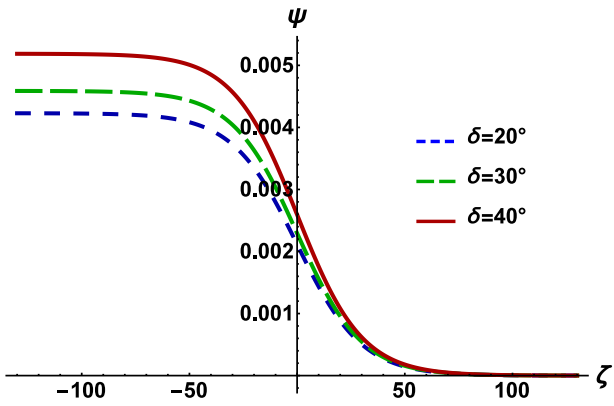}
\caption{The variation of positive potential $\psi$ with $\zeta$ for different values of $\delta$ when $\eta=0.3$,
$\kappa=1.6$, $\mu_e=0.5$, $\mu_p=0.3$, $\sigma_1=1.2$, $\sigma_2=0.2$, $\sigma_3=0.02$, $\sigma_4=1.5$, $U_o=0.01$, and $\Upsilon_p=\Upsilon_{p+}$.}
\label{1Fig:F2}
\end{figure}
\begin{figure}
\centering
\includegraphics[width=80mm]{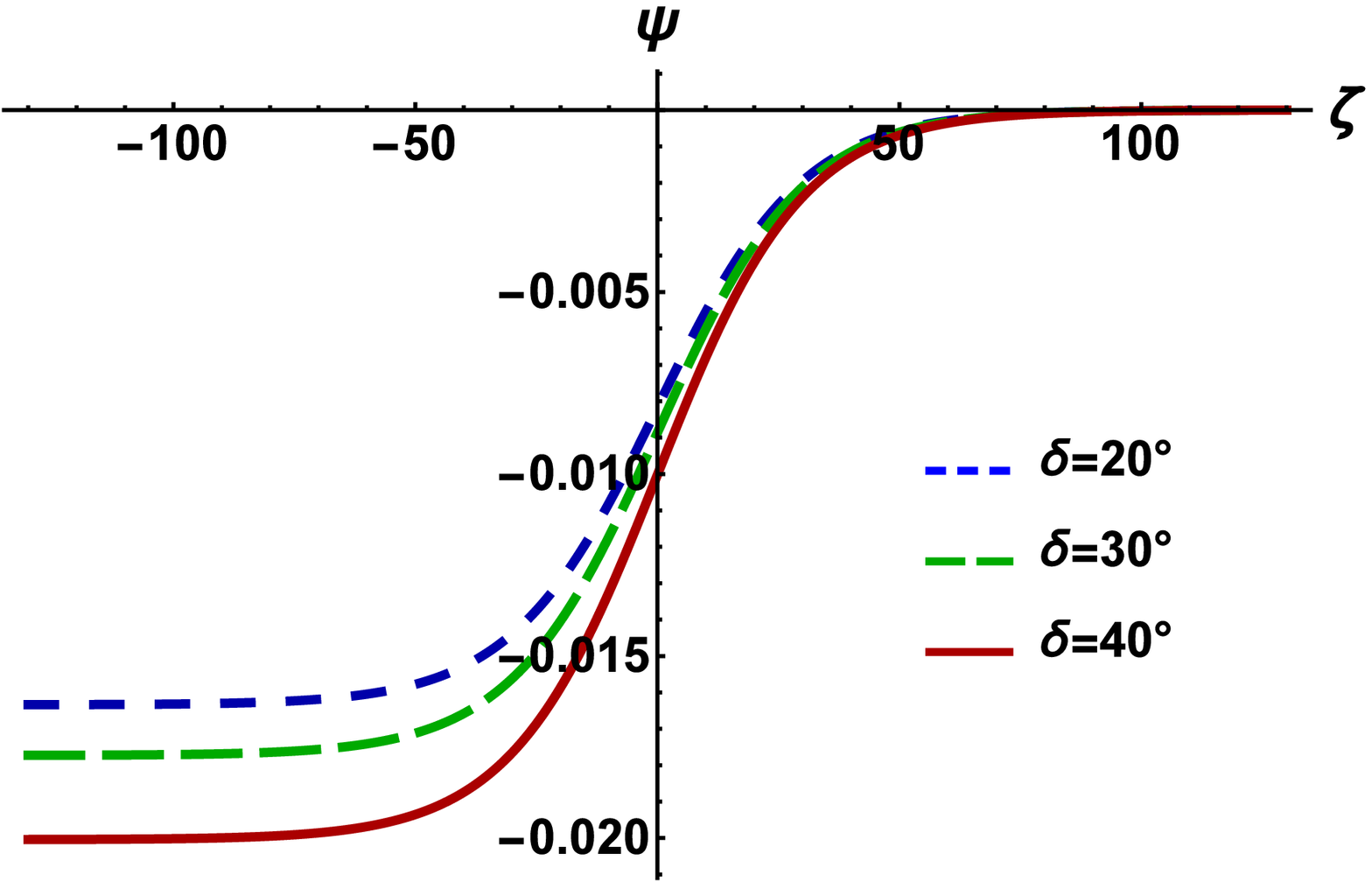}
\caption{The variation of negative potential $\psi$ with $\zeta$ for different values of $\delta$ when $\eta=0.3$,
$\kappa=1.6$, $\mu_e=0.5$, $\mu_p=0.3$, $\sigma_1=1.2$, $\sigma_2=0.2$, $\sigma_3=0.02$, $\sigma_4=0.5$, $U_o=0.01$, and $\Upsilon_p=\Upsilon_{p+}$.}
\label{1Fig:F3}
\end{figure}
\begin{figure}
\centering
\includegraphics[width=80mm]{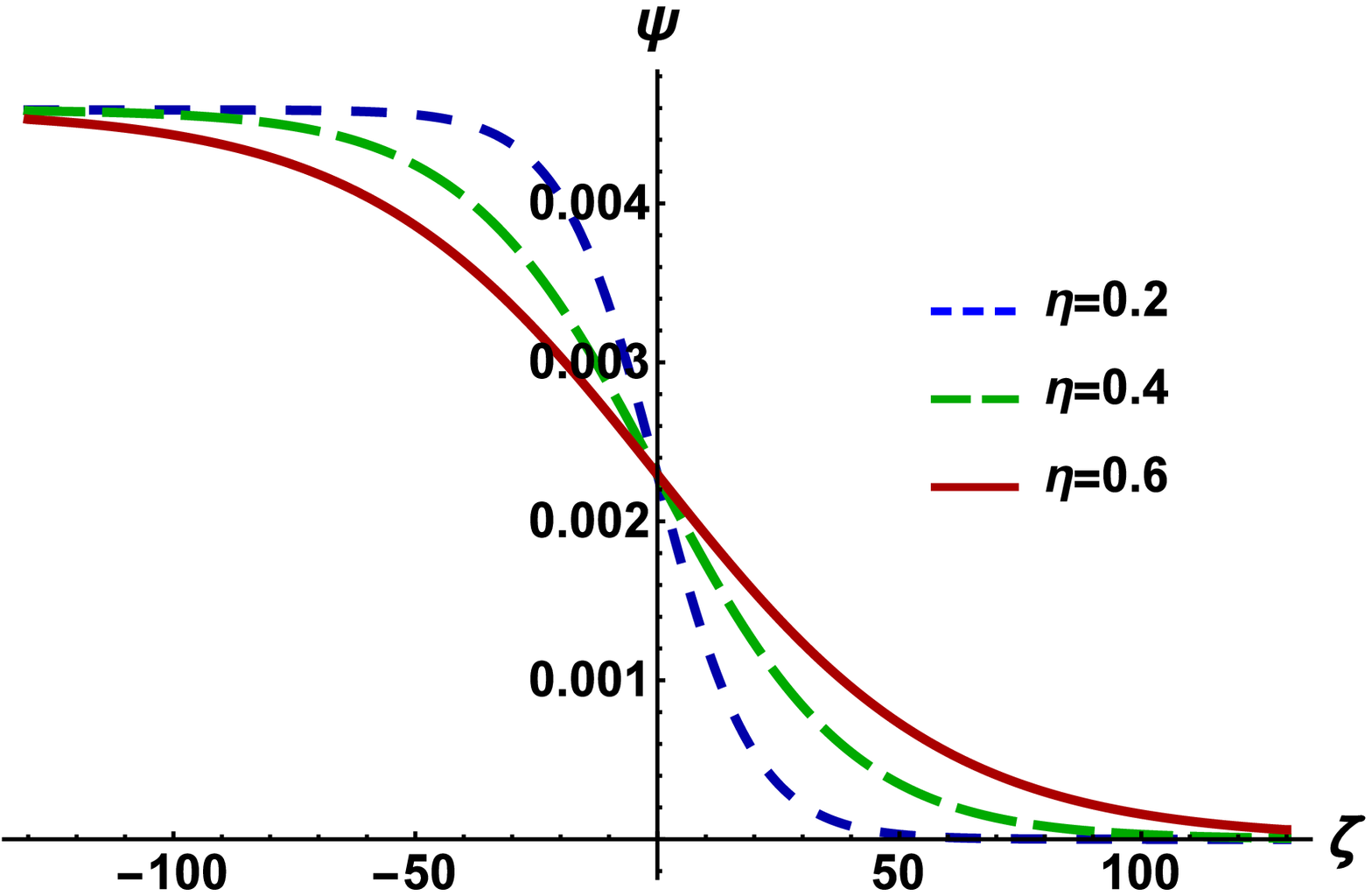}
\caption{The variation of positive potential $\psi$ with $\zeta$ for different values of $\eta$ when $\delta= 30^\circ$,
$\kappa=1.6$, $\mu_e=0.5$, $\mu_p=0.3$, $\sigma_1=1.2$, $\sigma_2=0.2$, $\sigma_3=0.02$, $\sigma_4=1.5$, $U_o=0.01$, and $\Upsilon_p=\Upsilon_{p+}$.}
\label{1Fig:F4}
\end{figure}
\begin{figure}
\centering
\includegraphics[width=80mm]{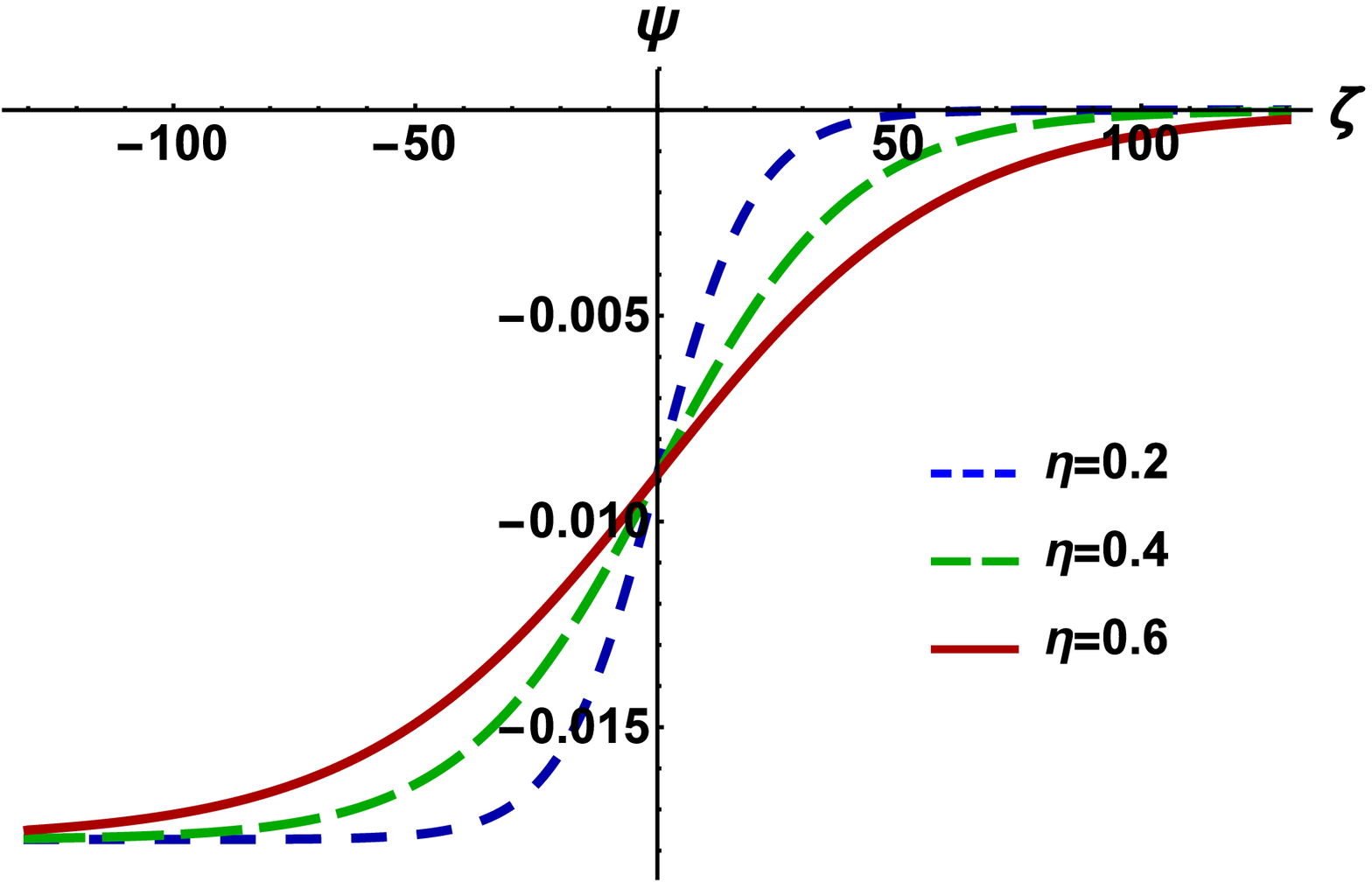}
\caption{The variation negative potential $\psi$ with $\zeta$  for different values of $\eta$  when $\delta= 30^\circ$,
$\kappa=1.6$, $\mu_e=0.5$, $\mu_p=0.3$, $\sigma_1=1.2$, $\sigma_2=0.2$, $\sigma_3=0.02$, $\sigma_4=0.5$, $U_o=0.01$, and $\Upsilon_p=\Upsilon_{p+}$.}
\label{1Fig:F5}
\end{figure}
\begin{figure}
\centering
\includegraphics[width=80mm]{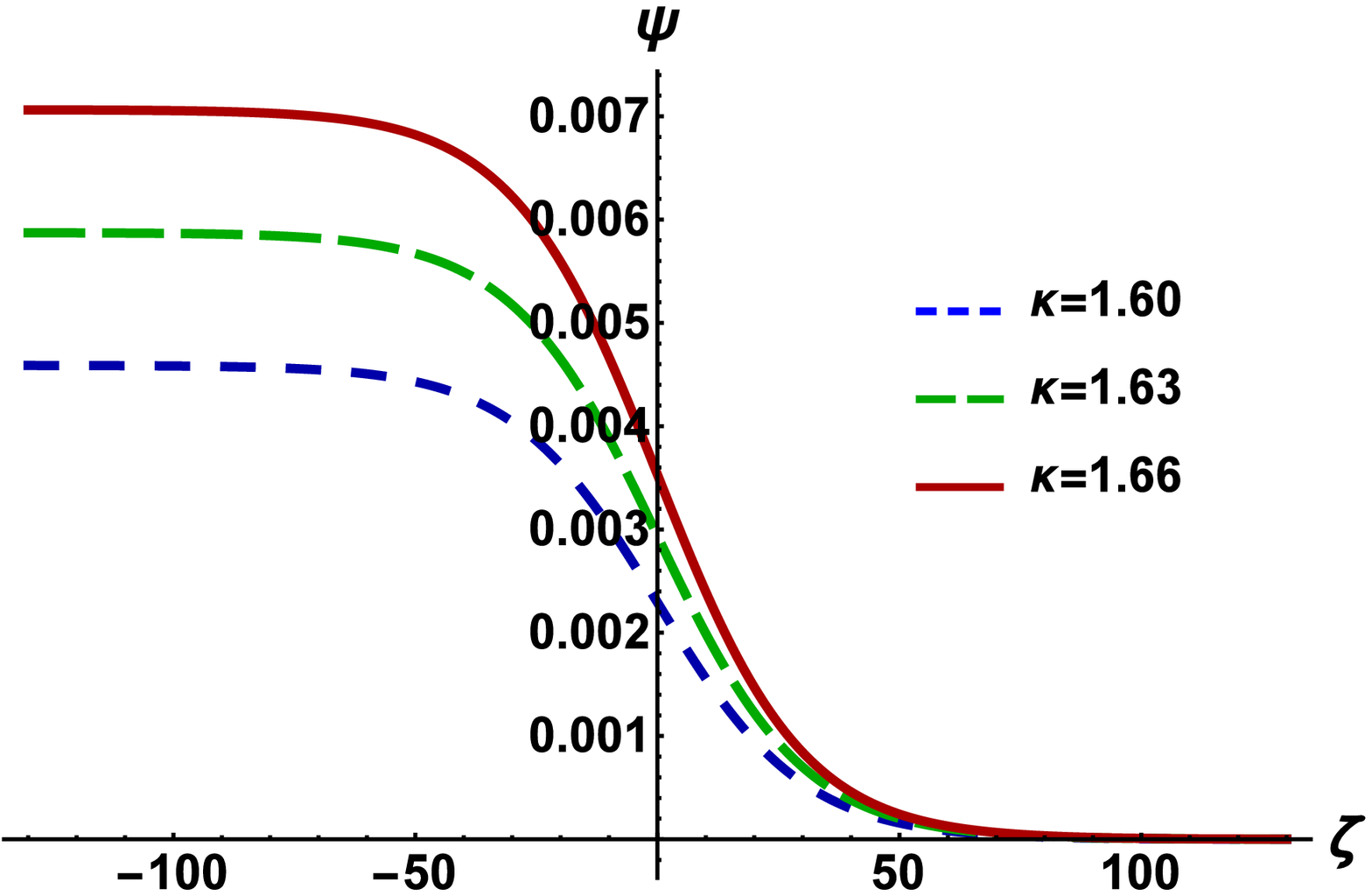}
\caption{The variation of positive potential $\psi$ with $\zeta$ for different values of $\kappa$ when $\delta= 30^\circ$, $\eta=0.3$,
 $\mu_e=0.5$, $\mu_p=0.3$, $\sigma_1=1.2$, $\sigma_2=0.2$, $\sigma_3=0.02$, $\sigma_4=1.5$, $U_o=0.01$, and $\Upsilon_p=\Upsilon_{p+}$.}
\label{1Fig:F6}
\end{figure}
\begin{figure}
\centering
\includegraphics[width=80mm]{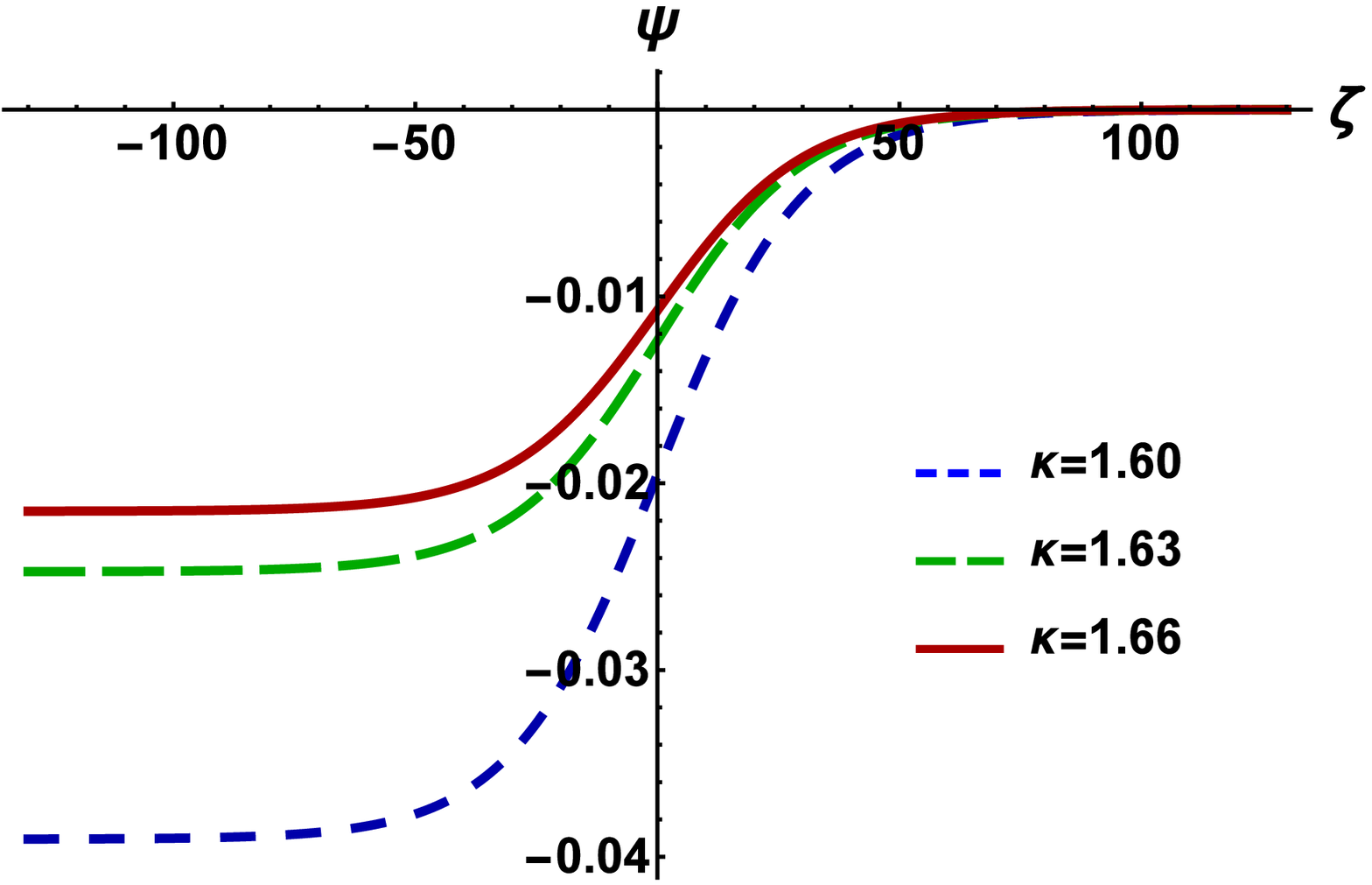}
\caption{The variation of negative potential $\psi$ with $\zeta$ for different values of $\kappa$ when $\delta= 30^\circ$, $\eta=0.3$,
$\mu_e=0.5$, $\mu_p=0.3$, $\sigma_1=1.2$, $\sigma_2=0.2$, $\sigma_3=0.02$, $\sigma_4=0.5$, $U_o=0.01$, and $\Upsilon_p=\Upsilon_{p+}$.}
 \label{1Fig:F7}
\end{figure}
\begin{figure}
\centering
\includegraphics[width=80mm]{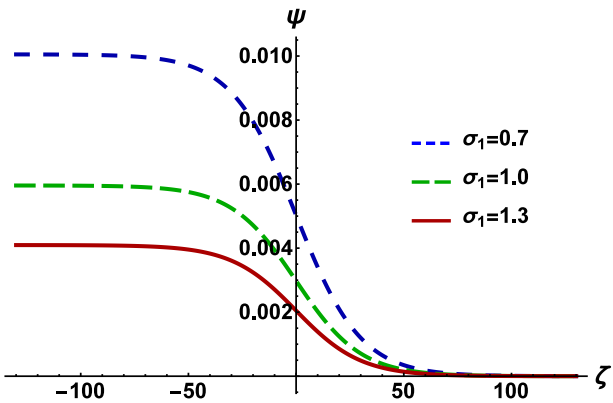}
\caption{The variation of positive potential with $\psi$ for different values of $\sigma_1$ when $\delta= 30^\circ$, $\eta=0.3$,
$\kappa=1.6$, $\mu_e=0.5$, $\mu_p=0.3$, $\sigma_2=0.2$, $\sigma_3=0.02$, $\sigma_4=1.5$, $U_o=0.01$, and $\Upsilon_p=\Upsilon_{p+}$.}
\label{1Fig:F8}
\end{figure}
\section{Results and discussion}
\label{1sec:Results and discussion}
The balance between nonlinearity and dissipation leads to generate IASHWs in a four-component
magnetized PIPM in the presence of super-thermal electrons and positrons. We have numerically analyzed the
variation of nonlinear coefficient ($A$) with $\sigma_4$ in Fig. \ref{1Fig:F1},
and it is obvious from this figure that initially $A$ is negative (i.e., $A<0$) for small
values of $\sigma_4$ ($\sigma_4<0.7$), and it becomes zero (i.e., $A=0$) for $\sigma_4\equiv0.7$. The value at
which $A$ becomes zero is known as critical value of $\sigma_4$ (i.e., $\sigma_{4c}$), and after the
critical value of $\sigma_4$, $A$ becomes positive (i.e., $A>0$) for large values of $\sigma_4$ ($\sigma_4>0.7$).
The parametric regime associated with $A<0$ and $\sigma_4<\sigma_{4c}$ allows to generate negative
electrostatic IASHWs (i.e., $\psi<0$) while the parametric regime associated with $A>0$ and $\sigma_4>\sigma_{4c}$
allows to generate positive IASHWs (i.e., $\psi>0$) when other plasma parameters are $\delta=30^\circ$,
$\eta=0.3$, $\kappa=1.6$, $\mu_e=0.5$, $\mu_p=0.3$, $\sigma_1=1.2$, $\sigma_2=0.2$, $\sigma_3=0.02$, and $\Upsilon_p=\Upsilon_{p+}$.
So, our plasma model allows to generate both positive and negative electrostatic shock structures according
to the values of $\sigma_4$.

The existing external magnetic field can significantly change the configuration of the IASHWs in PIPM,
and the effects of the external magnetic field can be observed from Figs. \ref{1Fig:F2} and \ref{1Fig:F3},
and it is clear from these figures that the magnitude of the amplitude of the positive (i.e., $\psi>0$) and
negative (i.e., $\psi<0$) electrostatic shock profiles increases with the increase in the value of oblique
angle $\delta$ which is the angle between the direction of the existing external magnetic field and the
direction of the propagation of the electrostatic wave, and this result agrees with the result of Hossen \textit{et al.} \cite{Hossen2017}.

The effects of the kinematic viscosity of ions on the formation of positive electrostatic potential (i.e., $\psi>0$)
associated with $A>0$ and $\sigma_4>\sigma_{4c}$, and the negative electrostatic potential (i.e., $\psi<0$)
associated with $A<0$ and $\sigma_4<\sigma_{4c}$ can be seen, respectively, in Figs. \ref{1Fig:F4} and \ref{1Fig:F5}.
The steepness of the positive and negative IASHWs decreases with the increase in the value of $\eta$ but the magnitude
of the amplitude remains invariant with the variation of $\eta$, and this result is similar with the result of Ref. \cite{Abdelwahed2016}.
So, the result from our present investigation is congruent with the previous work.

The shape of the electrostatic shock profiles in a magnetized PIPM in terms of the super-thermality of electrons and
positrons is presented in Figs. \ref{1Fig:F6} and \ref{1Fig:F7}. The super-thermality of electrons and positrons
not only changes the height of the electrostatic shock profiles but also causes to change the width of the
electrostatic shock profiles. The height of positive (negative) shock profile associated with $A>0$ and $\sigma_4>\sigma_{4c}$
($A<0$ and $\sigma_4<\sigma_{4c}$) increases with the increase in the value of $\kappa$.

The dynamics of the PIPM is rigourously changed by the variation of the mass and charge state of the
PI, and the variation of positive electrostatic shock profile (i.e., $\psi>0$) with $\sigma_1$ under consideration
of $A>0$ and $\sigma_4>\sigma_{4c}$ can be observed in Fig. \ref{1Fig:F8}. The amplitude of the positive electrostatic shock profile (i.e., $\psi>0$)
increases (decreases) with increasing  positive (negative) ion mass when their charge state remains constant. On the other hand,
as we increase the charge of the positive (negative) ion, the amplitude of the positive electrostatic shock profile decreases (increases).
So, the mass and charge state of the PI play an opposite role in the formation of the positive electrostatic shock profile
in a four-component PIPM in the presence of super-thermal electrons and positrons.
\section{Conclusion}
\label{1sec:Conclusion}
We have investigated IASHWs in a four-component magnetized PIPM by considering kinematic viscosities
of both inertial warm positive and negative ion species, and inertialess super-thermal electrons and positrons.
The reductive perturbation method is used to derive the Burgers' equation. The results that have been found from
our present investigation can be summarized as follows:
\begin{itemize}
  \item Both positive (i.e., $\psi>0$) and negative (i.e., $\psi<0$) electrostatic shock structures can be generated in PIPM
  having super-thermal electrons and positrons.
  \item The magnitude of the amplitude of positive and negative shock structures increases with oblique angle.
  \item The steepness of the shock profile decreases with the kinematic viscosity of ion.
  \item The height of the shock profile increases (decreases) with the mass of positive (negative) ion.
\end{itemize}
It may be noted here that the gravitational effect is needed to be considered in the governing equations but
beyond the scope of our present work.  The results of our present investigation will be useful
in understanding the nonlinear phenomena both in
astrophysical environments such as cometary comae \cite{Chaizy1991}, ($H^+$, $O_2^-$) and ($H^+$, $H^-$) plasmas in the D
and F-regions of Earth's ionosphere \cite{Massey1976}, upper regions of Titan's atmosphere \cite{Massey1976,Coates2007},
and also in the laboratory experiments, namely, plasma processing reactors \cite{Gottscho1986},  ($Ar^+$, $SF_6^-$) plasma \cite{Wong1975,Nakamura1997,Cooney1991,Nakamura1999}, ($Ar^+$, $O_2^-$) plasma, ($K^+$, $SF_6^-$) plasma \cite{Song1991,Sato1994},
($Xe^+$, $F^-$) plasma \cite{Ichiki2002}, neutral beam sources \cite{Bacal1979}, plasma etching \cite{Sheehan1988},
($Ar^+$, $F^-$) plasma \cite{Nakamura1984}, combustion products \cite{Sheehan1988},
and Fullerene ($C_{60}^+$, $C_{60}^-$) plasma \cite{Oohara2003,Oohara2005}, etc.

\end{document}